\documentclass[twocolumn,showpacs,prd]{revtex4}
\usepackage{epsfig}
\usepackage{bm}
\usepackage{amsfonts}
\usepackage{amssymb,amsmath}
\usepackage{enumerate}
\usepackage{multirow}
\usepackage{epsfig}
\usepackage{graphicx}
\usepackage{bm}
\begin{document}

\title{Quantum mechanical interaction of matter with the scalar mode of gravitational wave in modified gravity theories}
\author{Rakesh Das}
\email{rakeshkrishno@gmail.com}
\affiliation{Department of Physics, West Bengal State University, Barasat, North 24 Paraganas, West Bengal, India}
\author{Anirban Saha}
\email{anirban@wbsu.ac.in}
\affiliation{Department of Physics, West Bengal State University, Barasat, North 24 Paraganas, West Bengal, India}

\begin{abstract} 
We study the interaction of a quantum mechanical particle with gravitational wave (GW) in the framework of modified theory of gravity (MTG) where apart from the two standard tensorial modes of polarization of GW there exists another massless scalar mode.  The purpose of using the MTG framework in our study is to uncover key features in matter's response to GWs that,  if observed in actual GW data,  can serve as observational evidence in favor of MTG over standard General Relativity. 
\end{abstract}

\maketitle {}

\section{Introduction} 

 Since the discovery of the late-time accelerated expansion of the observable universe\cite{late1,late2} various attempts have been made to explain the source driving the said acceleration \cite{LCDM,  dark1}.  However,  the constraints on cosmological parameters put together by current observational data can not conclusively break the degeneracy \cite{dege_1,dege_2,dege_3,  dege_4} among the various theoretical models of the universe like the $\Lambda$CDM model \cite{LCDM}  a plethora of scalar field-driven accelerated Universe models \cite{dark1,  dark2,  dark3} also known as dark energy models, and scalar-tensor gravity or modified gravity models \cite{MTG_1, MTG_2}.    
Therefore apart from the continuous endeavour to develop stronger electromagnetic channels some alternative channels of observation are also being explored.  Since the first direct detection of gravitational waves (GW) in 2015 \cite{gra_detec1,gra_detec2}, GW astronomy,  which was hitherto inaccessible to us,  has opened up as a new data source for Astrophysical as well as Cosmological inference \cite{gra_data1, gra_data2,gra_data3,gra_data4,gra_data5,gra_data6}.  This not only provides us an observational probe into some strong-gravity scenarios like black-hole and neutron star mergers but also with their enhanced signal-to-noise ratio,  the upcoming ground-based GW observatories like the Einstein telescope \cite{Rdmp96},  the Cosmic  Explorer \cite{Rdmp97} and the space-based LISA mission\cite{Rdmp58}  can uncover signatures of new physics.  
For example,  the response of ordinary matter to GW signals may differ in the framework of standard General Relativity (GR) vis-\`a-vis modified theory of gravity (MTG).  So {\it a theoretical study of matter-GW interaction using the modified-gravity framework may uncover key features that,  if observed in actual GW data,  can serve as observational evidence in favor of MTG over standard GR. }  

Owing to the extremely small length-scale at which matter interacts with GW,  the interaction must be fundamentally quantum mechanical in nature.  Therefore,  in the present paper we propose to study the quantum mechanical interacton of a test particle with GW using the MTG framework in a four-dimensional spacetime as a simple first step to this end.  

The quantum mechanical interaction of matter with GW \cite{Weber, deWitt,  Speli} is worth persuing in its own right because of its fundamental nature.  Moreover it has a direct application in resonant detectors of GW \cite{maggiore,  NCGW1, NCGWSwarup}.  Naturally such a study in the MTG framework can provided valuable insight and predictions that can be confronted with experimental data,  especially since in this framework additional GW plarization modes show up \cite{fr_re_1,fr_re_2,fr_re_3,fr_re_4,fr_re_5}.


The organization of the paper is as follows: In the next section we explain how a quantum mechanical description of a simple matter system interacting with GWs can be obtained.  In section III we review the various polarization modes of GW in a modified theory of gravity framework and choose the suitables ones that serve the purpose of this paper.  In section IV we obtain the solution which shows the response of test matter to incoming GWs.  We discuss the results in section V.
\section{Methodology}
  
 Classically,  the dynamics of a test particle of mass $m$ interacting with GW,  is governed by the geodesic deviation equation which,  in the long wavelength and low-velocity limit\footnote{The long wavelength ($|x_{j}| << \frac{\lambda}{2\pi}$,  the reduced wavelength of GW) and low-velocity (velocities of the test mass are non-relativistic $v << c$) limit is where resonant bar-detectors and all the earth bound interferometric detectors of GW operates. },  takes the form~\cite{maggiore} 
\begin{equation}\label{geodesic}
 m\ddot{x}^{j}=-mR^{j}_{0,k0}x^{k}
\end{equation}
in a proper detector frame.
Here dot denotes derivative with respect to coordinate time of the proper detector frame,  $x^{j}$ is the proper distance of particle from the origin and 
\begin{eqnarray}
R^{j}_{0,k 0} = -\frac{1}{2}\ddot{h}_{jk}
\label{R_j0k0} 
\end{eqnarray}
are the components of the curvature tensor in terms of metric perturbation $h_{\mu\nu}$,  defined by decomposing the metric $g_{\mu\nu}$ as
\begin{align}\label{Eq:purturbation}
g_{\mu\nu}=\eta_{\mu\nu}+h_{\mu\nu};~~~|h_{\mu\nu}|<<1
\end{align}
in a linearised theory of gravity  $h_{\mu\nu}$ behaves as a second rank tensor under background Lorentz transformation.   $\eta_{\mu\nu}$ is the flat Minkowski background metric.  The metric perturbation $h_{\mu \nu}$ satisfies a wave equation and therefore is referred to as the gravitational wave.  

For a quantum mechanical description of this system we need to quantize an appropriate classical Hamiltonian,  which we can obtain from a suitable Lagrangian that would lead to the classical equation of motion given by Eqn.(\ref{geodesic}).  Upto a total time-derivative term that classical Lagrangian is 
\begin{align}\label{lagrangian}
\mathcal{L}=\frac{1}{2}m\dot{x}^{2}-m\Gamma^{j}_{0k} x_{j}x^{k}
\end{align}
where,  the affine connection  $\Gamma^{j}_{0k} $ is given by $\Gamma^{j}_{0k} =\dot{h}_{jk}/2$. 
Using the canonical momentum $p_{j}=m\dot{x}_{j} - m\Gamma^{j}_{0k}x^{k}$,  we can write the classical Hamiltonian of the system as
\begin{align}\label{hamilton}
\mathcal{H}=\frac{1}{2m}{\left( p_{j} + m\Gamma^{j}_{0k}x^{k}\right)}^2
\end{align}
which, due to the linearised nature of the theory,  simplifies to 
\begin{align}\label{hamilton_1}
\mathcal{H}=\frac{p_{j}^{2}}{2m}+ \Gamma^{j}_{0k} x_{j} p_{k}
\end{align}

To quantize this system we replace the canonical variables in (\ref{hamilton_1}) by the corresponding position and momentum operators and in terms of the raising and lowering operators $a_{j}$ and $a_{j}^{\dagger}$ defined by
\begin{align*}
\hat{x}_{j}=\left(\frac{\hbar}{2m\omega}\right)^\frac{1}{2}\left(a_{j}+a_{j}^{\dagger}\right);
\end{align*}
\begin{align}\label{operator}
\hat{p}_{j}=-i\left(\frac{m\omega\hbar }{2}\right)^{\frac{1}{2}}\left(a_{j}-a^{\dagger}\right)
\end{align}
the quantum mechanical Hamiltonian takes the form 
\begin{align}\label{hamilton_free}
\hat{\mathcal{H}}=\frac{\hbar\omega}{4}\left(2a_{j}a^{\dagger}_{j}+1-a^{\dagger 2}_{j}-a^{2}_{j}\right)-\frac{i\hbar}{4}\Gamma^{j}_{0k}\left(a_{k}a_{j}-a^{\dagger}_{k}a^{\dagger}_{j}\right)
\end{align}
Here the frequency $\omega$ is determined by the initial uncertainty in either the position or the momentum of the particle \cite{Speli}.
The Hamiltonian operator (\ref{hamilton_free}) drives the time evolution of the system. This can be depicted in terms of the Heisenberg equation of motion for $a_{j}(t)$
\begin{eqnarray}\label{operator_time}
\frac{da_{j}(t)}{dt}=\frac{i}{\hbar}\left[\mathcal{H},a_{j}\right] 
\end{eqnarray}
and that of $a^{\dagger}_{j}(t)$,  given by the complex conjugate of equation (\ref{operator_time}).

Using (\ref{hamilton_free}) and employing the algebra satisfied by the raising and lowering operators
\begin{eqnarray} \label{operator1}
\left[a_{j}(t),  a^{\dagger}_{k}(t)\right] &=&\delta_{jk} \nonumber\\
\left[a_{j}(t),a_{k}(t)\right] &=&0=
\left[a^{\dagger}_{j}(t),a^{\dagger}_{k}(t)\right] 
\end{eqnarray} 
the time-evolution equation (\ref{operator_time}) becomes 
\begin{eqnarray}\label{operator3}
\frac{da_{j}(t)}{dt} &=& -\frac{i\omega}{2}\left(a_{j}-a^{\dagger}_{j}\right)+\frac{\dot{h}_{jk}}{2}a_{k}
\end{eqnarray}
Introducing the time dependent Bogoliubov transformation 
\begin{eqnarray} \label{operator2}
a_{j}(t)&=& u_{jk}(t)a_{k}(0)+v_{jk}(t)a^{\dagger}_{k}(0) \nonumber\\
a^{\dagger}_{j}(t) &=& a^{\dagger}_{k}(0)\bar{u}_{kj}(t)+a_{k}(0)\bar{v}_{kj}(t) 
\end{eqnarray} 
which relates the operators $a_{j}(t)$~~and~~$a^{\dagger}_{j}(t)$ with their initial values at $t=0$,  the time evolution of the system can be cast in terms of the Bogoliubov coefficients $u_{jk}\left(t \right)$ and $v_{jk}\left(t \right)$ which are $3\times 3$ matrices.  Owing to eq. (\ref{operator1}),  the Bogoliubov coefficients must satisfy
\begin{align}\label{bogo1}
uv^{\top}=u^{\top} v;~~~~ uu^{\dagger}-vv^{\dagger}=\mathcal{I}
\end{align}
written in matrix form,  where T and $\dagger$ denote transpose and transpose of complex conjugate respectively.  $\mathcal{I}$ is the $3 \times 3$ identity matrix.\\
Corresponding to the initial values in eq. (\ref{operator2}) the appropriate boundary conditions for the Bogoliubov coefficients are 
\begin{eqnarray} \label{Bogo_bc}
u_{jk}(0)=\delta_{jk}\,, v_{jk}(0)=0
\end{eqnarray}
Using (\ref{operator2}),  the time-evolution in equation (\ref{operator3}) can be expressed in terms of a new pair of variables ($\zeta, \xi$) defined by 
\begin{align}\label{bogo2}
\zeta_{jk}=u_{jk}-v^{\dagger}_{jk};~~~\xi_{jk}=u_{jk}+v^{\dagger}_{jk}
\end{align}
as  
\begin{eqnarray}
\label{Eq:zeta_t}
\frac{d\zeta_{jk}}{dt}&=&-\frac{1}{2}\dot{h}_{jp}\zeta_{pk}\\
\label{Eq:xi_t}
\frac{d\xi_{jk}}{dt}&=&-i\omega\zeta_{jk}+\frac{1}{2}\dot{h}_{jp}\xi_{pk}
\end{eqnarray}
Solving this set of coupled differential equations we can obtain the quantum dynaimcs of the system that will describe the response of a test mass not only to the standerd plus and cross polarizations of the incoming GW,  but possibly also to the extra polarization modes that appears in a modified gravity scenario.  So before proceeding further we need to briefly elaborate on the possible polarization modes of GW that appears in the MTG framework.
\section{Polariztion modes in MTG}
In addition to the two tensor modes of polarization of GWs appearing in the standard GR framework,  GWs in a modified theory of gravity can have a maximum of four other polarization modes in four-dimensional spacetime.
Two of them are vector modes which we ignore in this work since most of the gravity theories in cosmologically viable scenarios,  do not have them \cite{Nishizawa09}.
The remaining two are scalar modes.  One of them,  known as the breathing mode,  displaces a circular arrangement of test particles whitin the transverse plane,  much like the two tensor modes of GW does in the standard GR framework.  But the other scalar mode,  known as the longitudinal one,  causes displacement of these test particles along the plane of porpagation of the GW.  
Evidently the scalar longitudinal mode can serve as a feature that distinguishes GW in MTG from that in standard GR.  

Therefore in the reminder of this paper we shall only consider the scalar longitudinal polarization along with the two standard tensorial modes,  known as plus and cross polarizations of GW.  Hence $h_{jk}$ can be most conveniently represented  by
\begin{align} \label{Eq:MGW}
h_{jk}(t)=\sum^{3}_{I=1} h_{I}\left(t \right)e^{I}_{jk}
\end{align}
where the polarization information is contained in the three independent $3\times3$ matrices~\cite{Nishizawa09}
\begin{eqnarray} \label{pol_basis}
e^{\times}_{jk} &=& e^{1}_{jk} =
\begin{pmatrix}
0 & 1 & 0 \\
1 & 0 & 0 \\
0 & 0 & 0
\end{pmatrix},  \nonumber\\
e^{+}_{jk} &=& e^{2}_{jk}=
\begin{pmatrix}
1 & ~0 & 0 \\
0 & -1 & 0 \\
0 & ~0 & 0
\end{pmatrix}, \nonumber \\
e^{{\rm s}}_{jk} &=& e^{3}_{jk}=\sqrt{2}
\begin{pmatrix}
0 & 0 & 0 \\
0 & 0 & 0 \\
0 & 0 & 1
\end{pmatrix}
\end{eqnarray}
and $ h_{\times}\left(t \right),  \, h_{+}\left(t \right),  \, h_{{\rm s}}\left(t \right)$ are the time-dependent\footnote{Because we are working in a long wavelength and low velocity limit,  in a plane-wave expansion of GW,  $h_{jk} = \int (A_{jk}e^{ikx} + A^{*}_{jk}e^{-ikx}) d^{3}k /(2\pi)^{3}$,  the spatial part $e^{-i\vec{k}. \vec{x}} \approx 1$ all over the detector site. Thus our concern is the time-dependent part of the GW only. 
For linearly polarized GWs the frequency $\Omega$ is contained in the time-dependent amplitude $h_{I}\left(t \right)$; e.g. ,  $h_{I}\left(t \right) = h_{I}\left(0\right) \cos \Omega t$.} amplitudes for $I = 1,2,3$.  Here, while choosing the polarization matrices (\ref{pol_basis}) we have assumed that the GW propagates along the $z-$direction,  thus $x-y$ plane is the transvers one.

\section{Solution}
With the form of $h_{jk}\left(t \right)$ in (\ref{Eq:MGW}) we need to solve the coupled differential equations (\ref{Eq:zeta_t}, \ref{Eq:xi_t}) for the matrices $\left( \zeta_{jk}, \xi_{jk}  \right)$.  
We expand them as 
\begin{eqnarray}
\zeta_{jk}(t)&=&\sum ^{9}_{M=1}A_{M}e^{M}_{jk} \label{Eq:zeta} \\
\xi_{jk}&=&\sum ^{9}_{M=1}B_{M}e^{M}_{jk} \label{Eq:xi}
\end{eqnarray}
Here $\left\{e^{M}{}_{jk}\right\}$ is a suitable basis for the space of $3 \times 3$ matrices,  which has,  along with $e^{1}_{jk}$,  $e^{2}_{jk}$ and $e^{3}_{jk}$ in (\ref{pol_basis}),  six other independent matrices

\begin{align*}
e^{4}_{jk}=
\begin{pmatrix}
0 & 0 & 1 \\
0 & 0 & 0 \\
1 & 0 & 0
\end{pmatrix}, \,\, 
e^{5}_{jk}=
\begin{pmatrix}
0 & 0 & 0 \\
0 & 0 & 1 \\
0 & 1 & 0
\end{pmatrix},
\end{align*}
\begin{align*}
e^{6}_{jk}=
\begin{pmatrix}
1 & 0 & 0 \\
0 & 1 & 0 \\
0 & 0 & 0
\end{pmatrix} , \,\, 
e^{7}_{jk}=
\begin{pmatrix}
~0 & 1 & 0 \\
-1 & 0 & 0 \\
~0 & 0 & 0
\end{pmatrix},
\end{align*}
\begin{align}
e^{8}_{jk}=
\begin{pmatrix}
~0 & 0 & 1 \\
~0 & 0 & 0 \\
-1 & 0 & 0
\end{pmatrix},
e^{9}_{jk}=
\begin{pmatrix}
0 & ~0 & 0 \\
0 & ~0 & 1 \\
0 & -1 & 0
\end{pmatrix}
\end{align}
This helps reduce the equations of motion (\ref{Eq:zeta_t},  \ref{Eq:xi_t}),  which are in the matrix form,  to a set of coupled first order differential equations in the $A_{M}$ and $B_{M}$ coefficients:
\begin{eqnarray}
\dot{A_{1}}&=&-\frac{1}{2}\left(A_{6}\dot{h}_{\times}+A_{7}\dot{h}_{+}\right)\nonumber\\
\dot{A_{2}}&=&\frac{1}{2}\left(A_{7}\dot{h}_{\times}-A_{6}\dot{h}_{+}\right)\nonumber\\
\dot{A_{3}}&=&-\frac{A_{3}\dot{h}_{\rm s}}{\sqrt{2}}\nonumber\\
\dot{A_{4}}&=&-\frac{\dot{h}_{\times}}{4}\left(A_{5}+A_{9}\right)-\frac{\dot{h}_{+}}{4}\left(A_{4}+A_{8}\right)-\frac{\dot{h}_{\rm s}}{2\sqrt{2}}\left(A_{4}-A_{8}\right)\nonumber\\
\dot{A_{5}}&=&-\frac{\dot{h}_{\times}}{4}\left(A_{4}+A_{8}\right)+\frac{\dot{h}_{+}}{4}\left(A_{5}+A_{9}\right)-\frac{\dot{h}_{\rm s}}{2\sqrt{2}}\left(A_{5}-A_{9}\right)\nonumber\\
\dot{A_{6}}&=&-\frac{1}{2}\left(A_{1}\dot{h}_{\times}+A_{2}\dot{h}_{+}\right)\nonumber\\
\dot{A_{7}}&=&\frac{1}{2}\left(A_{2}\dot{h}_{\times}-A_{1}\dot{h}_{+}\right)\nonumber\\
\dot{A_{8}}&=&-\frac{\dot{h}_{\times}}{4}\left(A_{5}+A_{9}\right)-\frac{\dot{h}_{+}}{4}\left(A_{4}+A_{8}\right)+\frac{\dot{h}_{\rm s}}{2\sqrt{2}}\left(A_{4}-A_{8}\right)\nonumber\\
\dot{A_{9}}&=&-\frac{\dot{h}_{\times}}{4}\left(A_{4}+A_{8}\right)+\frac{\dot{h}_{+}}{4}\left(A_{5}+A_{9}\right) \nonumber\\& & +\frac{\dot{h}_{\rm s}}{2\sqrt{2}}\left(A_{5}-A_{9}\right)
\end{eqnarray}
\begin{eqnarray}
\dot{B_{1}}&=&-i\omega A_{1}-\frac{1}{2}\left(B_{6}\dot{h}_{\times}+B_{7}\dot{h}_{+}\right)\nonumber\\
\dot{B_{2}}&=&-i\omega A_{2}+\frac{1}{2}\left(B_{7}\dot{h}_{\times}-B_{6}\dot{h}_{+}\right)\nonumber\\
\dot{B_{3}}&=&-i\omega A_{3}-\frac{B_{3}\dot{h}_{\rm s}}{\sqrt{2}}\nonumber\\
\dot{B_{4}}&=&-i\omega A_{4} -\frac{\dot{h}_{\times}}{4}\left(B_{5}+B_{9}\right)\nonumber\\& & -\frac{\dot{h}_{+}}{4}\left(B_{4}+B_{8}\right)-\frac{\dot{h}_{\rm s}}{2\sqrt{2}}\left(B_{4}-B_{8}\right)\nonumber\\
\dot{B_{5}}&=&-i\omega A_{5}-\frac{\dot{h}_{\times}}{4}\left(B_{4}+B_{8}\right)+\frac{\dot{h}_{+}}{4}\left(B_{5} +B_{9}\right)\nonumber\\& & -\frac{\dot{h}_{\rm s}}{2\sqrt{2}}\left(B_{5}-B_{9}\right)\nonumber\\
\dot{B_{6}}&=&-i\omega A_{6} -\frac{1}{2}\left(B_{1}\dot{h}_{\times}+B_{2}\dot{h}_{+}\right)\nonumber\\
\dot{B_{7}}&=&-i\omega A_{7}+\frac{1}{2}\left(B_{2}\dot{h}_{\times}-B_{1}\dot{h}_{+}\right)\nonumber\\
\dot{B_{8}}&=&-i\omega A_{8}-\frac{\dot{h}_{\times}}{4}\left(B_{5}+B_{9}\right)\nonumber\\& &  -\frac{\dot{h}_{+}}{4}\left(B_{4}+B_{8}\right)+\frac{\dot{h}_{\rm s}}{2\sqrt{2}}\left(B_{4}-B_{8}\right)\nonumber\\
\dot{B_{9}}&=&-i\omega A_{9}-\frac{\dot{h}_{\times}}{4}\left(B_{4}+B_{8}\right) \nonumber\\& & +\frac{\dot{h}_{+}}{4}\left(B_{5}+B_{9}\right)+\frac{\dot{h}_{\rm s}}{2\sqrt{2}}\left(B_{5}-B_{9}\right)
\end{eqnarray}
Noting from (\ref{Eq:purturbation}) that $|h_{I}\left(t \right)| << 1$,  these equations can be solved iteratively about their $h_{I}\left(t \right)=0$ solutions.
For a suitable boundary condition on $h_{I}$ we assume that the GW hits the system at t = 0 so that
\begin{align}\label{Eq:b.c}
h_{I}\left(t \right)=0~~(I=1,2,3),~~~ for~~~ t\leq 0
\end{align}
The first order solutions obtained by this iterative method are:
\begin{align}
A_{1}=-\frac{h_{\times}}{2},~~~
A_{2}=-\frac{h_{+}}{2},~~~
A_{3}=-\frac{h_{\rm s}}{2}+\frac{1}{\sqrt{2}},~~~
A_{6}=1 \nonumber \\
A_{4}=0; ~~~~~
A_{5}=0;~~~~~
A_{7}=0;~~~~~
A_{8}=0;~~~~~
A_{9}=0 \label{zero_order_sol_A}
\end{align}
And,
\begin{eqnarray}
B_{1}&=&\frac{i\omega h_{\times}t}{2}+\frac{i\omega}{2}\int t \dot{h}_{\times}dt -\frac{h_{\times}}{2}\nonumber\\
B_{2}&=&\frac{i\omega h_{+}t}{2}+\frac{i\omega}{2}\int t \dot{h}_{+}dt -\frac{h_{+}}{2}\nonumber\\
B_{3}&=&\frac{i\omega h_{\rm s}t}{2}-\frac{i\omega t}{\sqrt{2}}+\frac{i\omega}{2}\int t \dot{h}_{\times}dt -\frac{h_{\rm s}}{2}+\frac{1}{\sqrt{2}}\nonumber\\
B_{4}&=& 0;~~~
B_{5}=0;~~~
B_{6}=\left(1-i\omega t\right) \nonumber \\
B_{7}&=&0;~~~
B_{8}=0;~~~
B_{9}=0 \label{zero_order_sol_B}
\end{eqnarray}

Using (\ref{Eq:zeta}, \ref{Eq:xi}) we can reexpress the  solution (\ref{zero_order_sol_A},  \ref{zero_order_sol_B}) in terms of the matrix pair ($\zeta_{jk},\xi_{jk}$) which can be further substituted in (\ref{operator2}) via (\ref{bogo2}) to obtain the time-evolution of the raising and lowering operators $a_{j}(t)$ and $a^{\dagger}_{j}(t)$ in terms of their initial values $a_{j} (0)$ and $a^{\dagger}_{j}(0)$.  

From the definition of the raising and lowering operators (\ref{operator}) at initial time $t = 0$ we can relate $a_{j}(0)$ and $a^{\dagger}_{j}(0)$ with the initial position and momentum expectation values 
\begin{eqnarray}
\langle \hat{x}_{i}\left(0\right) \rangle & = & \left(X_{0}, Y_{0}, Z_{0} \right) \nonumber \\
\langle \hat{p}_{i}\left(0\right) \rangle & = & \left(P_{x0}, P_{y0}, P_{z0}\right)
\label{ini}
\end{eqnarray} 
of the quantum mechanical particle that is used as the test body.  Further,  the initial uncertainty in either the position or the momentum of this test body determines $\omega$.

Using the definition of the raising and lowering operators (\ref{operator}) at any arbitrary time,  now we can express the time evolution of the position and momentum expectation values $<\hat{x}_{i}(t)>$ and $<\hat{p}_{i}(t)>$ in terms of their initial values (\ref{ini})
\begin{eqnarray}
<\hat{x}_{1}(t)>&=& X_{0}\left[1-\frac{h_{+}}{2}\right]+ \frac{P_{x0}}{m}\left[ t - \frac{t h_{+}}{2} + \frac{1}{2} \int t \dot{h}_{+} dt\right]\nonumber\\& & -\frac{h_{\times}}{2}Y_{0}-\frac{P_{y0}}{2m}\left[th_{\times}+\int t \dot{h}_{\times} dt\right]\nonumber
\\
<\hat{x}_{2}(t)>&=& Y_{0}\left(1+\frac{h_{+}}{2}\right) + \frac{P_{y0}}{m}\left[t + \frac{th_{+}}{2}+\frac{1}{2}\int t \dot{h}_{+} dt\right] \nonumber\\ & & -\frac{h_{\times}}{2}X_{0} - \frac{P_{x0}}{2m}\left[th_{\times}-\int t \dot{h}_{\times} dt\right]  \nonumber
\\
<\hat{x}_{3}(t)>&=&\left[1-\frac{h_{\rm s}}{\sqrt{2}}\right]Z_{0} \nonumber\\& & + \frac{P_{z0}}{m}\left[t - \frac{th_{\rm s}}{\sqrt{2}} - \frac{1}{\sqrt{2}}\int t \dot{h}_{\rm s}dt\right]
\label{xsoln}
\end{eqnarray}
and,
\begin{eqnarray}
<\hat{p}_{1}(t)>&=& \left( 1-\frac{h_{+}}{2}\right) P_{x0}-\frac{h_{\times}}{2}P_{y0}\nonumber\\
<\hat{p}_{2}(t)>&=& \left( 1+\frac{h_{+}}{2}\right) P_{y0}-\frac{h_{\times}}{2}P_{x0}\nonumber\\
<\hat{p}_{3}(t)>&=& \left(1- \frac{h_{\rm s}}{\sqrt{2}}\right)P_{z0}
\label{psoln}
\end{eqnarray}
This is the formal solution that depicts the response of the test particle to incoming GW of linear polarization in an MTG framework.  How it differs from a similar test scenario in the framework of standard GR and wheather it can produce a consistant limit will be discussed in the next section. 
\section{Results  and Discussion }
We begine with the limiting scenario of no incoming GW.  This can be implimented by substituting $h_{I}\left(t\right) = 0, $ for $ I = \times,  +,  {\rm s}$ for all time $t$ in (\ref{xsoln}, \ref{psoln}).  This yields
\begin{eqnarray}
<\hat{x}_{1}(t)>&=& X_{0}+ \frac{P_{x0}}{m}t \nonumber\\
<\hat{x}_{2}(t)>&=& Y_{0}+\frac{P_{y0}}{m}t \nonumber\\
<\hat{x}_{3}(t)>&=& Z_{0}+\frac{P_{z0}t}{m} 
\end{eqnarray}
and,
\begin{eqnarray}
<\hat{p}_{1}(t)>&=&  P_{x0}\nonumber\\
<\hat{p}_{2}(t)>&=&  P_{y0}\nonumber\\
<\hat{p}_{3}(t)>&=&  P_{z0},
\end{eqnarray}
the general solution for a free particle with the chosen initial values of position and momenta (\ref{ini}).  

With the consistancy check in place we now compare our solution to that in a particle-GW interaction in standard GR framework.  There only the two tensor modes of polarization $\left( h_{\times},  \, h_{+} \right)$ are present.  Therefore the corresponding solution for a particle--GW interaction can be easily obtained from our result (\ref{xsoln}, \ref{psoln}) by switching off the scalar mode,  i.e.  by substituting $h_{s}\left( t \right) = 0$ for all $t$.  This gives
\begin{eqnarray}
<\hat{x}_{1}(t)>&=& X_{0}\left[1-\frac{h_{+}}{2}\right]+ \frac{P_{x0}}{m}\left[ t - \frac{t h_{+}}{2} + \frac{1}{2} \int t \dot{h}_{+} dt\right]\nonumber\\& & -\frac{h_{\times}}{2}Y_{0}-\frac{P_{y0}}{2m}\left[th_{\times}+\int t \dot{h}_{\times} dt\right]\nonumber\\
<\hat{x}_{2}(t)>&=& Y_{0}\left(1+\frac{h_{+}}{2}\right) + \frac{P_{y0}}{m}\left[t + \frac{th_{+}}{2}+\frac{1}{2}\int t \dot{h}_{+} dt\right] \nonumber\\ & & -\frac{h_{\times}}{2}X_{0} - \frac{P_{x0}}{2m}\left[th_{\times}-\int t \dot{h}_{\times} dt\right]  \nonumber\\
<\hat{x}_{3}(t)>&=& Z_{0}+\frac{P_{z0}t}{m} 
\end{eqnarray}
and,
\begin{eqnarray}
<\hat{p}_{1}(t)>&=& \left( 1-\frac{h_{+}}{2}\right) P_{x0}-\frac{h_{\times}}{2}P_{y0}\nonumber\\
<\hat{p}_{2}(t)>&=& \left( 1+\frac{h_{+}}{2}\right) P_{y0}-\frac{h_{\times}}{2}P_{x0}\nonumber\\
<\hat{p}_{3}(t)>&=& P_{z0}, 
\end{eqnarray}
which shows that in standard GR formalism the GW affects the particle's dynamics non-trivially only {\it in the transeverse plane}.
But in MTG framework,  due to the presence of the {\it additional scalar mode} $h_{{\rm s}}$ {\it GW also affects the particle's motion out of the transverse plane}.  This is a distinction that should have a clear observational footprint that,  if observed in actual GW data,  can searve as evidance in favour of Modified Theories of Gravity over the standard General Relativity.

	Note that the results obtained in this paper only formally demonstrates that the consideration of the particle-GW interaction in a modified gravity framework {\it does} produce non-trivial theoretical predictions which are distinctive from their standard GR counterpart.  However,  owing to the extremely small coupling of matter with GW in both the frameworks under consideration,  the position or momentum of the test body are not suitable observables in an actual GW detector.  Also,  realizing an observational setup for a quantum mechanical particle interacting with GW can be tricky.  Instead,  GW interacting with a quantum mechanical harmonic oscillator is realizable in resonant detectors of GWs,  where the resonant transitions of the phonon modes produced by the incoming GW act as the observational data.  This we plan to take up in a future communication.


\end{document}